# Anytime and Efficient Coalition Formation with Spatial and Temporal Constraints

Luca Capezzuto[0000−0003−4404−0998], Danesh Tarapore[0000−0002−3226−6861], and Sarvapali D. Ramchurn[0000−0001−9686−4302]

School of Electronics and Computer Science, University of Southampton, UK
{luca.capezzuto,d.s.tarapore,sdr1}@soton.ac.uk

**Abstract** The *Coalition Formation with Spatial and Temporal constraints Problem* (CFSTP) is a multi-agent task allocation problem where the agents are few and cooperative, the tasks are many, spatially distributed, with deadlines and workloads, and the objective is to find a schedule that maximises the number of completed tasks. The current state-of-the-art CFSTP solver, the *Coalition Formation with Look-Ahead* (CFLA) algorithm, has two main limitations. First, its time complexity is quadratic with the number of tasks and exponential with the number of agents. Second, as we show, its look-ahead technique is not effective in real-world scenarios, such as open multi-agent systems, where new tasks can appear at any time. Motivated by this, we propose an extension of CFLA, called *Coalition Formation with Improved Look-Ahead* (CFLA2). Since we cannot eliminate the limitations of CFLA in CFLA2, we also develop a novel algorithm to solve the CFSTP, the first to be anytime, efficient and with provable guarantees, called *Cluster-based Coalition Formation* (CCF). We empirically show that, in settings where the look-ahead technique is highly effective, CCF completes up to 30% (resp. 10%) more tasks than CFLA (resp. CFLA2) while being up to four orders of magnitude faster. Our results affirm CCF as the new state-of-the-art algorithm to solve the CFSTP.

## 1 Introduction

According to the Global Risks Report 2020 [25], natural and man-made disasters are in the top 5 risks in terms of likelihood and in the top 10 risks in terms of impact. The reason is that they are strictly related to other important issues, such as: extreme weather events; biodiversity loss and ecosystem collapse; water and food crises; failure of climate-change mitigation and adaptation; failure of regional or global governance, and profound social instability. Consequently, the response phase plays an important role in the modern disaster management cycle [1,5].

In the field of *Multi-Agent Systems* (MASs), one of the most important projects promoting research on disaster response is the RoboCup rescue simulation [14].

By reproducing the aftermath of an earthquake in a city, this simulation allows testing coordination approaches that could be enacted by first responders in such situations. In this work, we are interested in a class of task allocation problems that can be generated by the RoboCup rescue simulation, namely, those in which ambulances have to find and rescue victims trapped under rubble, and fire brigades have to extinguish fires. This class of problems has been characterised by Ramchurn et al. [23] as *Coalition Formation with Spatial and Temporal constraints Problem* (CFSTP)[1]. In the CFSTP, agents (e.g., ambulances or fire brigades) have to decide which sequences of tasks (e.g., victims or fires) they are going to execute (e.g., save or extinguish). Their decision is influenced by how tasks are located in the disaster area, how much time it is required to reach them, how much work they require (e.g., how large a fire is) and their deadlines (e.g., estimated time left before victims perish). Given these conditions, and considering that there could be many more tasks than agents, it is crucial that agents cooperate with each other by forming coalitions [26] (i.e., grouping together). Hence, the objective of the CFSTP is to schedule the right coalitions (e.g., ambulances with the largest capability) to the right tasks (e.g., sites with the most victims) to ensure that as many tasks as possible are completed.

In this paper, our interest is in algorithms that solve the CFSTP *efficiently* and with *provable guarantees* (i.e., approximation algorithms [21]) and are *anytime* (i.e., which can return partial solutions if they are interrupted before completion). The reason is that being efficient and anytime is a desirable feature of real-world applications [32]. To date, approaches based on the Max-Sum algorithm [9] have proven to be among the most effective at solving the CFSTP, as well as many other problems [10]. The variants relevant to our scope are *Fast Max-Sum* (FMS) [23] and *Binary Max-Sum* (BinaryMS) [22]. FMS is anytime and provides optimal solutions in exponential time, but it cannot solve general CFSTP instances. On the other hand, BinaryMS is efficient, but not anytime, and it requires a pre-processing phase with exponential run-time to solve general CFSTP instances. Multi-agent approaches that solve problems similar to the CFSTP make use of social insects [8], automated negotiation [11,12,30] and evolutionary computation [31], but without considering the anytime property. In the iTax taxonomy of Korsah et al. [17], the CFSTP is defined as *Cross-schedule Dependent Single-Task Multi-Robot Time-extended Assignment* (XD [ST-MR-TA]) [17]. To date, the approaches proposed to solve XD [ST-MR-TA] problems utilise linear programming [2,15,16], automated negotiation [18] and memetic algorithms [19]. However, either they do not produce anytime solutions [18,19], or do not have provable guarantees [2,15], or are based on a simpler model [16].

Against this background, we focus on the current state-of-the-art algorithm that solves the CFSTP, namely, the *Coalition Formation with Look-Ahead* (CFLA) algorithm [24]. Our rationale is that CFLA is anytime and, even though its computational time is exponential in the worst case, thanks to its design [24, Section 6] and the performance of current computers, a well-engineered implementation can solve problems with hundreds of agents and thousands of tasks in minutes. We study it thoroughly and produce the following contributions:

– A CFSTP model and a CFLA definition that are clearer, more concise and more detailed than [24].
– CFLA2, an enhancement of CFLA.
– CCF, the first anytime and approximation algorithm to solve the CFSTP, which asymptotically outperforms both CFLA and CFLA2.

---

[1] We use the definitions of *coalition* and *coalition formation* given by [13,24,26]. Hence, a coalition is a flat and task-oriented organisation of agents, short-lived and dissolved when no longer needed, while coalition formation is a consequence of the emergent behaviour of the MAS [20].

The rest of the paper is organised as follows. In Section 2, we give our CFSTP model. Section 3 is dedicated to CFLA2. Given that CFLA2 keeps the core limitations of CFLA, Section 4 presents the novel CCF algorithm. Section 5 reports our empirical evaluation, and Section 6 concludes.

## 2 The CFSTP model

In this section, we first give our terminology, then characterise coalition allocations and values, and finally give the constraints and objective function of the CFSTP.

### 2.1 Basic definitions

Let $V = \{v_1, \ldots, v_m\}$ be a set of $m$ tasks and $A = \{a_1, \ldots, a_n\}$ be a set of $n$ agents[2]. Let $L_V$ and $L_A$ be respectively the set of all possible task and agent locations, not necessarily disjoint. Hence, more than one agent or task can be at the same location. Time $t$ is discrete, that is, $t \in \mathbb{N}$, each problem starts at $t = 0$ and agents travel or execute tasks in measurable time units. The time units needed by an agent to travel from one location to another are given by $\rho : A \times (L_A \cup L_V) \times L_V \to \mathbb{N}$. Unlike [24], we put $A$ in the domain of $\rho$ to characterise agents with different speeds[3]. Task locations do not change over time, while agent locations can. Each task $v$ has a *demand* $D_v = \{w_v, d_v\}$, where $w_v \in \mathbb{R}^+$ is the *workload* of $v$, or the amount of work required to complete $v$, and $d_v \in \mathbb{N}$ is the *deadline* of $v$, or the time until which agents can work on $v$. Our notion of work will be clear in Section 2.3. Hence, workloads can only be positive, and some tasks may have a deadline of zero[4].

We denote the location of agent $a$ at time $t$ by $l_a^t \in L_A \cup L_V$, the times at which $a$ starts and finishes working on task $v$ by $s_a^v \in [0, d_v]$ and $f_a^v \in [s_a^v, d_v]$, respectively, and the *latest deadline* by $d_{max} = \max_{v \in V} d_v$.

---

[2] Although not necessary, it is typically assumed that $m \gg n$.

[3] In real-world scenarios, this avoids approximating different speeds to the same one.

[4] In other words, a problem may have tasks that cannot be completed in time, independently of the algorithm chosen to solve it.

## 2.2    Coalition allocations

Agents are cooperative [29] and can work together to complete a task. Let $Part(A)$ be the set of partitions of $A$. A subset of agents $C \in Part(A)$ is called a *coalition*. At time $t$, the rationale for allocating coalition $C$ to task $v$ is that $C$ can complete $v$ in the lowest time possible. An *agent allocation* is denoted by $\tau_t^{a \to v}$ and represents the fact that agent $a$ works on task $v$ at time $t$. The *set of all agent allocations* is denoted by:

$$T = \{\tau_t^{a \to v}\}_{a \in A, \, v \in V, \, t \in [0, \, d_{max}]} \tag{1}$$

and contains all the combinatorially different agent allocations. A *coalition allocation* is denoted by $\tau_t^{C \to v}$ and represents the fact that coalition $C$ works on task $v$ at time $t$. Given a set of agent allocations $T' \subseteq T$, and a time $t' \le d_{max}$, the *set of coalition allocations corresponding to $T'$* over the time period $[0, t']$ is denoted by:

$$\Gamma(T', t') = \left\{\tau_t^{C \to v} \mid C = \{a \mid \tau_t^{a \to v} \in T'\}, t \le t'\right\} \tag{2}$$

Furthermore, the *set of all coalition allocations* is denoted by:

$$\Gamma = \Gamma(T, d_{max}) \tag{3}$$

Similar to $T$, $\Gamma$ contains all the combinatorially different coalition allocations. An agent allocation $\tau_t^{a \to v}$ is also denoted as a *singleton coalition allocation* $\tau_t^{\{a\} \to v}$.

## 2.3    Coalition values

Each coalition allocation has a *coalition value*, given by the function[5] $u : Part(A) \times V \to \mathbb{R}^+$. Unlike [24], we put $V$ in the domain of $u$ to characterise the fact that the same coalition may execute different tasks with different performances. Hence, given a coalition allocation $\tau_t^{C \to v}$, the value $u(C, v)$ expresses the amount of work that coalition $C$ does on task $v$ at each time $t$. The workload $w_v$ decreases linearly over time, depending only on $u(C, v)$.

## 2.4    Constraints

There are three constraint types: structural, temporal and spatial. Structural constraints require that each task $v$ can be allocated to only one coalition at a time. This is characterised by the following sets:

$$\forall v \in V, \Gamma_v = \left\{\Gamma' \subseteq \Gamma : \tau_t^{C_1 \to v}, \tau_t^{C_2 \to v} \in \Gamma' \implies C_1 = C_2\right\} \tag{4}$$

With an abuse of notation, we write $\tau_t^{C \to v} \in \Gamma_v$ to indicate that $\tau_t^{C \to v}$ belongs to a not specified set of $\Gamma_v$.

---

[5] In cooperative game theory, this is a *characteristic function* [3, Section 2.1].

Temporal constraints require that each task $v$ can be completed only within its deadline $d_v$. This is characterised by the function $\Delta : V \times \Gamma \to \{0, 1\}$, defined as follows:

$$\Delta(v, \Gamma) = \begin{cases} 1, & \text{if } \exists\, t \le d_v : \sum_{t' \le t,\ \tau_{t'}^{C \to v} \in \Gamma_v} u(C, v) \ge w_v \\ 0, & \text{otherwise} \end{cases} \tag{5}$$

Equation 5 utilises $\Gamma_v$ (Equation 4) to count only well-formed coalition allocations.

Spatial constraints require that an agent will not start working on a task before reaching it. This is characterised as follows:

$$\forall a \in A,\ \forall v \in V, \forall t \le d_v,\ s_a^v \ge t + \rho(a, l_a^t, l_v) \tag{6}$$

$$\forall a \in A, \forall v_1, v_2 \in V,\ f_a^{v_1} + \rho(a, l_{v_1}, l_{v_2}) \le s_a^{v_2} \tag{7}$$

A set of agent allocations $T' \subseteq T$ is called *legal* if it exists a time $t' \le d_{max}$ such that $\Gamma(T', t')$ satisfies Equation 5. A set of coalition allocations $\Gamma' \subseteq \Gamma$ that satisfies Equations 5, 6 and 7 is called *feasible*. Consequently, at time $t$, if $\tau_t^{C_1 \to v_1}$ and $\tau_t^{C_2 \to v_2}$ are feasible coalition allocations and $l_{v_1} \ne l_{v_2}$, then $C_1 \cap C_2 = \emptyset$.

### 2.5 Objective function

The objective function of the CFSTP is to find a feasible set of coalition allocations that maximises the number of completed tasks. More formally:

$$\arg\max_{\Gamma' \subseteq \Gamma} \sum_{v \in V} \Delta(v, \Gamma') \text{ subject to Equations 6 and 7} \tag{8}$$

Since, for each agent, we may need to consider all the possible agent allocations until $d_{max}$, the time complexity of Equation 8 is $O(|A| \cdot |V|! \cdot (d_{max})^{|V|})$.

A feasible set of coalition allocations $\Gamma' \subseteq \Gamma$ is called a *solution with degree $k$* if $\sum_{v \in V} \Delta(v, \Gamma') = k$, with $0 < k \le |V|$. Moreover, $\Gamma'$ is called a *partial solution* if $k \le |V|$ and an *optimal solution*[6] if $k = |V|$. Hence, the argument of the maxima in Equation 8 is a solution with the highest degree.

Ramchurn et al. [24] proved that the CFSTP is NP-hard [21], and a generalisation of the Team Orienteering Problem [4], which is a generalisation of the Travelling Salesman Problem [28]. As we said in Section 1, CFLA is the current state-of-the-art CFSTP solver. In the next section, we show how it can be improved.

## 3 Coalition Formation with improved Look-Ahead

We now present the *Coalition Formation with improved Look-Ahead* (CFLA2), an extension of the CFLA algorithm [24]. More precisely, its look-ahead phase

---

[6] Optimal solutions may not exist (see Footnote 4 at Page 3).

---

**Algorithm 1:** getLegalAgentAllocations (Phase 1 of CFLA2)

---

**Input:** time $t$

**Output:** the set of legal agent allocations at time $t$

1  $L_t \leftarrow \emptyset$

2  **for** $a \in A^t_{free}$ **do** // for each free agent $a$

3      **for** $v \in V_{unc}$ **do** // for each uncompleted task $v$

4          **if** $t + \rho(a, l^t_a, l_v) \le d_v$ **then** // if $a$ can reach $v$ at $t$ within $d_v$

5              $L_t \leftarrow L_t \cup \{\tau^{a \to v}_{t'}\}_{t + \rho(a, l^t_a, l_v) \le t' \le d_v}$

---

(Section 3.4) has two modifications that, as we shall see in Section 5, enhance the overall performance.

The concept of CFLA2 is the same as CFLA, but for completeness we briefly report it in Section 3.1. After that, we detail the procedures that compose CFLA2, explaining how they differ from the ones of CFLA. Finally, we list the limitations that CFLA2 continues to keep from CFLA, which are the rationale for our new algorithm in Section 4.

CFLA and CFLA2 have the same four phases, but [24] describes them in three algorithms. For readability purposes, we describe them in four algorithms.

### 3.1 The concept of CFLA2

CFLA2 is a centralised, anytime and greedy algorithm that solves Equation 8 by maximising the working time of the agents and minimising the time required by coalitions to complete tasks. It is divided into four phases:

1. Defining the legal agent allocations (Section 3.2).
2. For each task $v$, choosing the best coalition $C$ (Section 3.3).
3. For each task $v$, doing a 1-step look-ahead (Section 3.4) to define its *degree* $\delta_v$, or the number of tasks that can be completed after the completion of $v$.
4. At each time $t \in [0, d_{max}]$, allocating a task not yet completed and with the highest degree (Section 3.5).

We detail them below.

### 3.2 Phase 1: defining the legal agent allocations

At time $t$, Algorithm 1 determines which free agents[7] ($A^t_{free}$) can reach which uncompleted tasks ($V_{unc}$) before their deadlines. The resulting set of legal agent allocations is denoted by $L_t$. This phase is identical in CFLA.

### 3.3 Phase 2: selecting the best coalition for each task

Given a task $v$ and a set of legal agent allocations $L_t$ (computed by Algorithm 1), Algorithm 2 returns the *Earliest-Completion-First* (ECF)[8] coalition $C^*_v$ that can

---

[7] That is, agents who neither are travelling to nor working on a task.

[8] This logic is adapted from the *Earliest-Deadline-First* (EDF) scheduling [27].

---

**Algorithm 2:** ECF (Phase 2 of CFLA2)

---

**Input:** task $v$, a set of legal agent allocations $L_t$
**Output:** ECF coalition $C$

1   $A_v^t \leftarrow$ define from $L_t$ the agents that can reach $v$ at $t$ within $d_v$
2   $C_v^* \leftarrow \emptyset$ // the ECF coalition
3   $t_v^* \leftarrow d_v + 1$ // time at which $C_v^*$ completes $v$
4   $i \leftarrow 1$
5   **while** $i \leq |A_v^t|$ *and* $C_v^* = \emptyset$ **do**
6    **for** $C \in$ *all combinations of $i$ agents in $A_v^t$* **do**
7     **if** $\sum_{\tau_{t'}^{C \to v} \in \Gamma_v,\ C' \subseteq C,\ t' \in [t, d_v]} u(C, v) \geq w_v$ **then**
8      $t_{minmax} \leftarrow \min_{t_{max}} \left( w_v - \sum_{\tau_{t'}^{C \to v} \in \Gamma_v,\ C' \subseteq C,\ t' \in [t, t_{max}]} u(C, v) \right)$
9      **if** $t_{minmax} < t_v^*$ **then**
10       $t_v^* \leftarrow t_{minmax}$
11       $C_v^* \leftarrow C$
12   $i \leftarrow i + 1$

---

be allocated to $v$. More precisely, the algorithm minimises both the size of $C_v^*$ and the time at which it completes $v$. This is achieved by iterating from the smallest to the largest possible coalition size (line 5) and iterating through all the possible coalitions of each size (line 6). When the procedure finds a coalition $C$ that can complete $v$ within its deadline (line 7), then $|C|$ is the minimum size of the coalitions that can complete $v$. Hence, $C_v^*$ is identified among the coalitions that have size $|C|$ (lines $8 - 11$).

Algorithm 2 is more concise than the original formulation [24, Algorithm 2]. In particular, we clarify that the minimum coalition size has to be determined by iterating through the subsets of the combinations[9] of $A_v^t$, which is the set of free agents that at time $t$ can reach $v$ within $d_v$.

### 3.4 Phase 3: defining the degree of each task

Given a task $v$, Algorithm 3 does a 1-step look-ahead[10] to define its degree $\delta_v$ (Section 3.1). Similarly to Algorithm 2, it checks how many tasks can be completed after the completion of $v$ (line 8).

Algorithm 3 differs from the original look-ahead phase [24, Algorithm 3] in two points. First, it only considers uncompleted tasks that have a deadline greater or equal to $d_v$ (line 4): this prevents from counting tasks that can be completed before the completion of $v$. In fact, as defined in Section 3.1, $\delta_v$ represents the number of tasks that can be completed only after the completion of $v$, not also those that are completed before that. Second, at line 11, $\delta_v$ is

---

[9] To date, the most efficient technique to enumerate all such combinations is the Gray binary code [7, Section 7.2.1.1].
[10] Which can be seen as a brute force phase.

---

**Algorithm 3:** lookAhead (Phase 3 of CFLA2)

---

**Input:** task $v$, its ECF coalition $C_v^*$, the set of all agent allocations $T$
**Output:** the degree $\delta_v$ of task $v$

1  $\delta_v \leftarrow 0$
2  $f_v \leftarrow$ time at which $C_v^*$ completes $v$
3  **for** $v_2 \in V_{unc} \setminus \{v\}$ **do**
4      **if** $d_{v_2} \geq d_v$ **then**
5          $A_{free}^{f_v} \leftarrow$ agents that are free at $f_v$ `// derived from` $C_v^*$ `and T`
6          $A^{d_{v_2}} \leftarrow$ select from $A_{free}^{f_v}$ the agents that can reach $v_2$ within $d_{v_2}$
7          $i \leftarrow 1$
8          **while** $i \leq |A^{d_{v_2}}|$ **do**
9              **for** $C \in$ *all combinations of $i$ agents in* $A^{d_{v_2}}$ **do**
                `// if` $C$ `can complete` $v_2$
10                 **if** $\sum_{\tau_t^{C' \to v} \in \Gamma_v,\ C' \subseteq C,\ t \in [f_v, d_{v_2}]} u(C, v) \geq w_v$ **then**
11                     $\delta_v \leftarrow \delta_v + 1 + (1 - \eta_{v_2})$
12                     $i \leftarrow |A^{d_{v_2}}|$ `// break external loop too`
13                     **break**
14             $i \leftarrow i + 1$

---

not just incremented by 1, but also by $1 - \eta_{v_2}$, where $\eta_{v_2}$ is the normalisation of $w_{v_2}$ in the interval $[w_{min}, w_{max}]$, with $w_{min}$ and $w_{max}$ being respectively the minimum and maximum task workloads. Hence, $\delta_v$ is also a measure of how much total workload is left after the completion of $v$. When $\delta_v$ is maximised (line 12 of Algorithm 4) it leads to the remaining tasks with the smallest workloads, thus increasing the probability of completing more.

### 3.5   Phase 4: overall procedure of CFLA2

Algorithm 4 shows the overall procedure. It runs in iterations until all tasks are completed or the latest deadline is expired. At each time $t$, it updates the set of legal agent allocations (line 8). Then, it determines which task to allocate to which coalition (lines $9-18$). If no other tasks can be allocated, the algorithm stops early (line 19).

### 3.6   Analysis and discussion

Algorithm 1 iterates through all free agents and uncompleted tasks. Assuming that line 4 requires constant time, the time complexity is $\alpha = O(|A| \cdot |V|)$.

Algorithm 2 iterates (line 5) from coalition size 1 to $|A_v^t|$, where $A_v^t$ is the set of agents that can reach task $v$ at time $t$. This requires $O(|A|)$ time. For each $s \leq |A_v^t|$, all possible coalitions of size $s$ could be examined (line 6), which requires $O(2^{|A|})$ time in case $A_v^t = A$. Assuming that line 8 requires $O(d_{max})$ time, the total time complexity is $\beta = O(|A| \cdot 2^{|A|} \cdot d_{max})$.

---

**Algorithm 4:** Overall procedure (Phase 4 of CFLA2)

---

**1** $t \leftarrow 0$
**2** $T \leftarrow \{\tau_t^{a \rightarrow v}\}_{a \in A, v \in V, t \in [0, d_{max}]}$ `// the set of all agent allocations`
**3** $V_{unc} \leftarrow V$ `// uncompleted tasks`
**4 repeat**
**5**     $\delta_{max} \leftarrow 0$ `// maximum task degree`
**6**     $v^* \leftarrow \text{NIL}$ `// next task to allocate`
**7**     $C^* \leftarrow \emptyset$ `// coalition to which` $v^*$ `is allocated`
**8**     $L_t \leftarrow \text{getLegalAgentAllocations}(t)$ `// Algorithm 1`
**9**     **for** $v \in V_{unc}$ **do**
**10**       $C_v^* \leftarrow \text{ECF}(v, L_t)$ `// Algorithm 2`
**11**       $\delta_v \leftarrow \text{lookAhead}(v, C_v^*, T)$ `// Algorithm 3`
**12**       **if** $\delta_v > \delta_{max}$ **then**
**13**          $\delta_{max} \leftarrow \delta_v$
**14**          $C^* \leftarrow C_v^*$
**15**     **if** $v^* \neq \text{NIL}$ *and* $C^* \neq \emptyset$ **then**
**16**       Allocate $C^*$ to $v^*$
**17**       $V_{unc} \leftarrow V_{unc} \setminus \{v^*\}$
**18**       Reduce $T$ according to new agent locations and availability
**19**     **if** $A_{free}^t = A$ **then** `// all agents are free`
**20**       **break**
**21**     $t \leftarrow t + 1$
**22 until** $V_{unc} = \emptyset$ *or* $t > d_{max}$

---

Algorithm 3 iterates through all uncompleted tasks, which requires $O(|V|)$ time, and its loop at line 8 is computationally identical to line 5 in Algorithm 2. Hence, the time complexity is $\gamma = O(|V| \cdot 2^{|A|})$.

Since it uses the previous algorithms, Algorithm 4 has a time complexity of

$$O\left(d_{max} \cdot (\alpha + |V| \cdot (\beta + \gamma))\right) = O\left((d_{max} \cdot |V|)^2 \cdot 2^{|A|}\right) \qquad (9)$$

Therefore, despite having a lower complexity than an optimal CFSTP solver (Section 2.5), CFLA2 has a run-time that increases quadratically with the number of tasks and exponentially with the number of agents. This makes the algorithm not suitable for systems with limited computational power or real-time applications. Other limitations are as follows:

1. It can allocate only one task per time [24, Section 7]. More formally, at each time, if one or more tasks are allocable, the worst- and best- case guarantee of CFLA2 is to find a partial solution with degree $k = 1$.
2. In general, greedily allocating a task with the highest degree now does not ensure that uncompleted tasks can all be successfully allocated in future. This is particularly relevant in an open system, where there is no certainty of having further uncompleted tasks (Section 1).

3. The more the tasks can be grouped by degree, the more the look-ahead phase becomes a costly random choice. In other words, at time $t$, if some tasks $V^{'} \subseteq V$ have all maximum degree, then Algorithm 4 selects $v^*$ randomly from $V^{'}$. Hence, the larger $V^{'}$ is, the less relevant Algorithm 3 becomes.

4. In Algorithm 4, all tasks have the same weight. That is, tasks with earlier deadlines may not be allocated before tasks with later deadlines. This is independent of the order in which the uncompleted tasks are elaborated (line 9). In fact, the computation of $\delta_{max}$ (line 12) would not be affected.

These limitations prevent CFLA2 from scoring higher percentages of completed tasks. Because of them, the next section presents a CFSTP solver that is anytime, efficient and with provable guarantees.

## 4 Cluster-based Coalition Formation

The *Cluster-based Coalition Formation* (CCF) is an anytime and greedy algorithm that operates at the agent level, rather than at the coalition level. It is divided into two phases:

1. For each agent $a$, defining the closest and most urgent uncompleted task that can be allocated to $a$.
2. For each task $v$, defining the minimum coalition of agents to which $v$ has to be allocated.

Algorithm 5 is used in Phase 1, while Algorithm 6 enacts the two phases. We describe them respectively in Sections 4.1 and 4.2.

### 4.1 Selecting the best task for each agent

Given a time $t$ and an agent $a$, Algorithm 5 returns the uncompleted task $v$ that is allocable, the most urgent and closest to $a$. By *allocable* we mean that $a$ can reach $v$ before deadline $d_v$, while *most urgent* means that $v$ has the earliest deadline. The algorithm prioritises unallocated tasks, that is, it first tries to find a task to which no agents are travelling, and on which no agents are working ($v_a^t[0]$). Otherwise, it returns an already allocated but still uncompleted task such that $a$ can reach it and contribute to its execution ($v_a^t[1]$). This ensures that an agent becomes free only when no other tasks are allocable and uncompleted.

Algorithm 5 does not enforce constraints on the workloads. As we shall see in Section 4.2, it is Algorithm 6 that does it, by allocating a task $v$ to a coalition $C$ only when $C$ has the minimum size and can complete $v$ within $d_v$.

### 4.2 Overall procedure of CCF

The overall procedure is described in Algorithm 6. The `repeat-until` structure is the same as CFLA2, to preserve the anytime property. Phases 1 and 2 are represented respectively by the loops at lines 5 and 16.

---

**Algorithm 5:** getTaskAllocableToAgent (used in Phase 1 of CCF)

---

**Input:** time $t$, agent $a$

**1** $v_a^t \leftarrow$ (NIL, NIL) // **array of indices 0 and 1**

**2** $t_{min} \leftarrow (d_{max}+1, d_{max}+1)$ // **like above**

**3** $d_{min} \leftarrow (d_{max}+1, d_{max}+1)$ // **like above**

**4** **for** $v \in V$ **do** // **for each uncompleted task**

**5**      $i \leftarrow 0$ // $v$ **is unallocated**

**6**      **if** *other agents are travelling to or working on $v$* **then**

**7**          $i \leftarrow 1$ // $v$ **is allocated but still uncompleted**

**8**      $t_{arr} \leftarrow t + \rho(a, l_a^t, l_v)$

**9**      **if** $t_{arr} \leq d_v$ *and* $t_{arr} < t_{min}[i]$ *and* $d_v < d_{min}[i]$ **then**

**10**          $v_a^t[i] \leftarrow v$

**11**          $t_{min}[i] \leftarrow t_{arr}$

**12**          $d_{min}[i] \leftarrow d_v$

**13** **if** $v_a^t[0] \neq$ NIL **then** // **prioritise unallocated tasks**

**14**      **return** $v_a^t[0]$

**15** **return** $v_a^t[1]$

---

Phase 1 loops through all agents. Here, an agent $a$ may either be free or reaching a task location. In the first case (line 6), if an uncompleted task $v$ can be allocated to $a$ (lines $7-8$), then $v$ is flagged as allocable (line 9) and $a$ is added to the set of agents $A_v^t$ to which $v$ could be allocated at time $t$ (line 11). In the second case (line 12), $a$ is travelling to a task $v$, hence its location is updated (line 13) and, if it reached $v$, it is set to *working on $v$* (line 14).

Phase 2 visits each uncompleted task $v$. If $v$ is allocable (line 18), then it is allocated to the smallest coalition of agents in $A_v^t$ (defined in Phase 1) that can complete it (lines $19-32$). In particular, at lines $24-27$, $\varphi_v$ is the amount of workload $w_v$ done by all the coalitions formed during the arrival to $v$ of the first $i-1$ agents in $\Pi_v^t$ (defined at line 19). After that, if there are agents working on $v$ (line 33), its workload $w_v$ is decreased accordingly (line 34). If $w_v$ drops to zero or below, then $v$ is completed (lines $35-37$). The algorithm stops (line 39) when all the tasks have been completed, or the latest deadline is expired, or no other tasks are allocable and uncompleted (Section 4.1).

### 4.3 Analysis and discussion

The approach of CCF transforms the CFSTP from a 1-$k$ task allocation to a series of 1-1 task allocations. In other words, instead of allocating each task to a coalition of $k$ agents, we have that coalitions are formed by clustering (i.e., grouping) agents based on the closest and most urgent tasks. Algorithm 5 runs in $\psi = O(|V|)$ time, assuming that the operation at line 8 has constant time. In Algorithm 6, the time complexity of Phase 1 is $O(|A| \cdot \psi) = O(|A| \cdot |V|)$, while Phase 2 runs in $O(|V| \cdot |A| \log |A|)$ because: in the worst case, $A_v^t = A$ and line 19 sorts $A$ in $\Omega(|A| \cdot \log |A|)$ time using any comparison sort algorithm [6]; the

---

**Algorithm 6:** Overall procedure of CCF (Phases 1 and 2)

---

**Input:** tasks $V$, agents $A$, task locations $L_V$, initial agent locations $L_A$, task demands $\{D_v\}_{v \in V}$

**Output:** A set of coalition allocations $\Gamma'$

1   $t \leftarrow 0$
2   $\Gamma' \leftarrow \emptyset$ `// the partial solution to return`
3   $V_{allocable} \leftarrow \emptyset$ `// allocable tasks`
4   **repeat**
5     **for** $a \in A$ **do** `// Phase 1`
6       **if** $a \in A^t_{free}$ **then**
7         $v \leftarrow$ getTaskAllocableToAgent$(t, a)$ `// Algorithm 5`
8         **if** $v \neq$ *NIL* **then**
9           **if** $v \notin V_{allocable}$ **then**
10             $V_{allocable} \leftarrow V_{allocable} \cup \{v\}$
11           $A^t_v \leftarrow A^t_v \cup \{a\}$
12       **else**
13         Update $a$'s location
14         **if** *a reached the task v it was assigned to* **then**
15           Set $a$'s status to *working on v*
16     **for** $v \in V$ **do** `// Phase 2`
17       $C^t_v \leftarrow$ all agents working on $v$ at time $t$
18       **if** $v \in V_{allocable}$ **then**
19         $\Pi^t_v \leftarrow$ list of all agents in $A^t_v$ sorted by arrival time to $v$
20         $C^* \leftarrow \emptyset$
21         **for** $i \leftarrow 1$ **to** $|\Pi^t_v|$ **do**
22           $C^* \leftarrow$ first $i$ agents in $\Pi^t_v$
23           $\lambda_i \leftarrow$ arrival time to $v$ of the $i$-th agent in $\Pi^t_v$
24           **if** $i + 1 \leq |\Pi^t_v|$ **then**
25             $\lambda_{i+1} \leftarrow$ arrival time to $v$ of the $(i+1)$-th agent in $\Pi^t_v$
26           **else**
27             $\lambda_{i+1} \leftarrow d_v$
28           $\varphi_v \leftarrow \varphi_v + (\lambda_i + \lambda_{i+1}) \cdot u(C^* \cup C^t_v, v)$ `// w_v done at` $\lambda_{i+1}$
29           **if** $(d_v - \lambda_i) \cdot u(C^*, v) \geq w_v - \varphi_v$ **then**
30             **break** `// C* is the minimum coalition to complete` $v$
31         $T_v = \bigcup_{a \in C^*} \{\tau^{a \rightarrow v}_{\lambda_a}\}$ `// λ_a is a's arrival time to` $v$
32         $\Gamma' \leftarrow \Gamma' \cup \Gamma(T_v, t)$ `// add` $\Gamma(T_v, t)$ `(Section 2.2) to` $\Gamma'$
33         $V_{allocable} \leftarrow V_{allocable} \setminus \{v\}$
34       **if** $C^t_v \neq \emptyset$ **then**
35         $w_v \leftarrow w_v - u(C^t_v, v)$
36         **if** $w_v \leq 0$ **then**
37           Set free all agents in $C^t_v$
38           $V \leftarrow V \setminus \{v\}$
39     $t \leftarrow t + 1$
40   **until** $V = \emptyset$ *or* $t > d_{max}$ *or all agents are free*

loop at line 21 runs in $O(|A|)$ time. Since the `repeat-until` structure is executed at most $d_{max}$ times, the time complexity of Algorithm 6 is:

$$O\left(d_{max} \cdot |V| \cdot |A| \log |A|\right) \tag{10}$$

If both phases are executed in parallel, the time complexity is reduced to:

$$\Omega\left(d_{max} \cdot (|V| + |A| \log |A|)\right) \tag{11}$$

CCF does not have the limitations of CFLA2 because:

1. It can allocate at least one task per time. More formally, at each time, if one or more tasks are allocable, CCF guarantees to find a partial solution with degree $1 \le k \le |A|$.
2. Each agent is always assigned to the allocable task that is closest and with the earliest deadline.
3. It runs in polynomial time and does not have a look-ahead phase. Thus, it is efficient and can be used in open systems.

**Theorem 1.** *CCF is correct.*

*Proof.* We prove by induction on time $t$.

At $t = 0$, a task $v$ is selected for each agent $a$ such that $v$ is allocable, the most urgent and closest to $a$ (Section 4.1). This implies that the agent allocation $\tau_0^{a \to v}$ is legal (Section 2.4). Then, Phase 2 of Algorithm 6 (Section 4.2) allocates $v$ to $a$ only if it exists a coalition $C$ such that $|C|$ is minimum, $\tau_0^{C \to v}$ is feasible (Section 2.4) and $a \in C$.

At $t > 0$, for each agent $a$, there are two possible cases: a task $v$ has been allocated to $a$ at a time $t' < t$, or $a$ is free (i.e., idle). In the first case, $a$ is either reaching or working on $v$ (lines $12 - 15$ in Algorithm 6), hence $\tau_t^{a \to v}$ is legal and $\tau_t^{C \to v}$ is feasible, where $a \in C$. In the second case, $a$ is either at its initial location or at the location of a task on which it finished working at time $t' < t$. Thus, as in the base case, if it exists a coalition $C$ and a task $v$ such that $|C|$ is minimum, $\tau_t^{C \to v}$ is feasible and $a \in C$, then $v$ is allocated to $a$.

As we said above, CCF can allocate between 1 and $|A|$ tasks at each time. However, its greedy approach does not allow to define the degree of the partial solution it converges to, regardless of the problem instance. Since no current algorithm that solves the CFSTP is simultaneously anytime, efficient and with provable guarantees (Section 1), CCF is the first of its kind.

## 5   Empirical evaluation

We implemented CFLA, CFLA2 and CCF in Java[11], and replicated the experimental setup of [24] because we wanted to evaluate how well CFLA2 and CCF perform in settings where the look-ahead technique is highly effective. For each test configuration, we solved 100 random CFSTP instances and plotted the average and standard deviation of: percentage of completed tasks; agent travel time[12]; *task completion time*, or the time at which a task has no workload left; *problem completion time*, or the time at which no other tasks can be allocated.

---

[11] https://git.soton.ac.uk/cmi/gopal/cfstp

### 5.1  Setup

Let $U(l, u)$ and $U^I(l, u)$ be respectively a uniform real distribution and a uniform integer distribution with lower bound $l$ and upper bond $u$. Our parameters are defined as follows:

- All agents have the same speed.
- The initial agent locations are randomly chosen on a 50 by 50 grid, where the travel time between two points is given by the Manhattan distance[13].
- Tasks are fixed to 300, while agents range from 2 to 40, in intervals of 2 between 2 and 20 agents, and in intervals of 5 between 20 and 40 agents.
- The coalition values are defined as $u(C, v) = |C| \cdot k$, where $k \in U(1, 2)$. Hence, coalition values depend only on the number of agents involved, and all tasks have the same difficulty.
- Deadlines $d_v \in U^I(5, 600)$ and workloads $w_v \in U^I(10, 50)$.

Unlike [24], we set the number of maximum agents to 40, instead of 20, because it allows, in this setup, to complete all tasks in some instances.

### 5.2  Results

In terms of completed tasks (Figure 1a), the best performing algorithm for instances with up to 18 agents is CFLA2, while the best performing algorithm for instances with at least 20 agents is CCF. CFLA is outperformed by CFLA2 in all instances except those with 2 agents, and by CCF in instances with at least 10 agents. The reason why the performance of CFLA and CFLA2 does not improve significantly starting from instances with 20 agents is that the more agents (with random initial locations) there are, the more tasks are likely to be grouped by degree[14]. CFLA2 has a similar trend to CFLA because it has the same limitations, but it performs better thanks to its improved look-ahead phase.

Regarding agent travel times (Figure 1b), it can be seen that CCF is up to three times faster than CFLA and CFLA2. This is due to Algorithm 5, which allocates tasks to agents also based on their proximity. To explain why agent travel times increase with all algorithms[15], let us consider a toy problem with one agent $a_1$ and one task $v$. If we introduce a new agent $a_2$ such that $\rho(a_2, l_{a_2}^0, l_v) > \rho(a_1, l_{a_1}^0, l_v)$, then the average travel time increases. In our scenario, this happens because the initial locations of the agents are random.

---

[12] See Section 2.1.
[13] Also known as taxicab metric or $\ell_1$ norm.
[14] See Limitation 3 described in Section 3.6.
[15] This behaviour is also reported, but not explained, in [24].

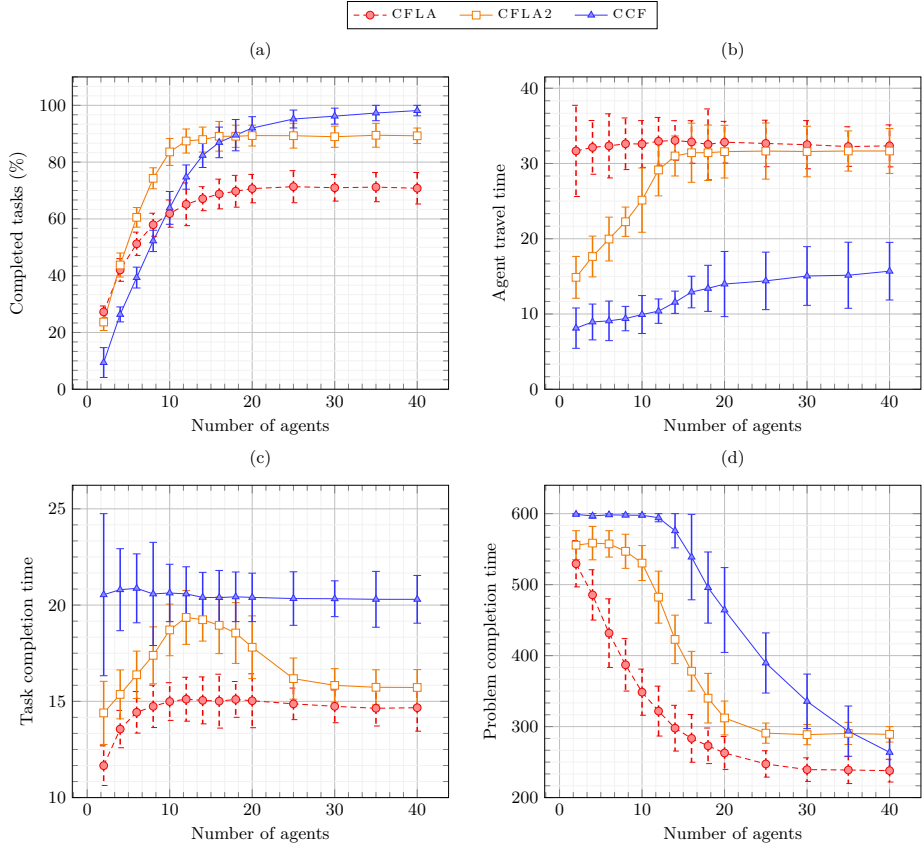

**Figure 1.** Tests on instances with 300 tasks and up to 40 agents.

In general, task completion times (Figure 1c) decrease because the more agents there are, the faster the tasks are completed. The completion of task $v$ is related to the size of the coalition $C$ to which $v$ is allocated: the highest the completion time, the smallest the size of $C$, hence the highest the working time of the agents in $C$. Task completion times are inversely related to agent travel times. Since CCF has the smallest agent travel times and allocates tasks to the smallest coalitions, it consequently has the highest task completion times. Therefore, in CCF, agents work the highest amount of times, and the number of tasks attempted at any one time is the greatest.

The problem completion times (Figure 1d) are in line with the task completion times (Figure 1c) since the faster the tasks are completed, the less time is needed to solve the problem. The reason why the times of CFLA and CFLA2 do not decrease significantly from 20 agents up is linked to their performance (see the discussion on Figure 1a above). On the other hand, the fact that the times of CCF decrease more consistently than those of CFLA and CFLA2 indicates that CCF is the most efficient asymptotically. In other words, CCF is likely to solve large-scale problems in fewer time units than CFLA and CFLA2.

In terms of computational times, CCF is significantly faster than CFLA and CFLA2. For example, in instances with 40 agents and 300 tasks, on average[16] CCF is $45106\% \pm [2625, 32019]$ (resp. $27160\% \pm [1615, 20980]$) faster than CFLA (resp. CFLA2). The run-time improvement of CFLA2 is due to line 4 of Algorithm 3, thanks to which the look-ahead phase elaborates fewer tasks.

## 6    Conclusions

In this paper, we proposed two novel algorithms to solve the CFSTP. The first is CFLA2, an improved version of the heuristic CFLA, and the second is CCF, which is the first to be anytime, efficient and with provable guarantees. CFLA2 can be used in place of CFLA for small and static problems, while CCF provides a baseline for benchmarks with dynamic and large-scale problems. Given that it significantly outperforms CFLA and is more applicable than CFLA2, we can consider CCF to be the new state-of-the-art algorithm to solve the CFSTP.

The limitation of CCF is that it cannot define the quality of of its approximation (Section 4.3). In particular, the fact that it maximises the agent working times (Section 5) implies that some agents may take longer to complete some tasks and therefore might not work on others. Thus, if an optimal solution exists, CCF cannot guarantee to obtain it.

Future work aims at developing the first anytime and optimal algorithm to solve the CFSTP. We also want to create distributed versions of CCF and our future algorithm, to define a large-scale benchmark from real-world datasets and to test on hard problems generated with the RoboCup rescue simulation.

## Acknowledgments

We thank Mohammad Divband Soorati, Ryan Beal and the anonymous reviewers for their helpful comments and suggestions. This research is sponsored by the AXA Research Fund. Danesh Tarapore acknowledges support from a EPSRC New Investigator Award grant (EP/R030073/1).

---

[16] On a machine with an Intel Core i5-4690 processor (quad-core 3.5 GHz, no hyper-threading) and 8 GB DDR3-1600 RAM.